# On the Origin of Plasticity-Induced Microstructure Change under Sliding Contacts

Yilun Xu[1,2,*], Daniel S. Balint[1], Christian Greiner[3,4], Daniele Dini[1*]

[1]Department of Mechanical Engineering, Imperial College London, London SW7 2AZ, UK
[2]Department of Materials, Imperial College London, London SW7 2AZ, UK
[3]Karlsruhe Institute of Technology (KIT), Institute for Applied Materials (IAM), Kaiserstrasse 12, 76131 Karlsruhe, Germany
[4]KIT IAM-CMS MicroTribology Center (µTC), Strasse am Forum 5, 76131 Karlsruhe, Germany

**Abstract**

Discrete dislocation plasticity (DDP) calculations are carried out to investigate the response of a single crystal contacted by a rigid sinusoidal asperity under sliding loading conditions to look for causes of microstructure change in the dislocation structure. The mechanistic driver is identified as the development of lattice rotations and stored energy in the subsurface, which can be quantitatively correlated to recent tribological experimental observations. Maps of surface slip initiation and substrate permanent deformation obtained from DDP calculations for varying contact size and normal load suggest ways of optimally tailoring the interface and microstructural material properties for various frictional loads.

**Key Words**: Discrete dislocation plasticity; Sliding; Size effect; Microstructure change

[*] Corresponding author: yilun.xu@imperial.ac.uk; d.dini@imperial.ac.uk, Tel./Fax: +442075947242.

# 1  Introduction

The resistance of a material surface to the interaction with another contacting surface as they slide reflects the performance of the material to the application of loads resulting in complex stress fields whose extent, magnitude and effects on the development of permanent microstructural changes vary depending on many factors. The presence of roughness due to asperities of practical engineering surfaces gives rises that the actual contact area usually takes only a small fraction of the nominal contact area [1]. Hence, the contact size governs the material response to tribological loads when it comes to the length scale of single asperity and grain size [2].

Frictional sliding is a complicated phenomenon generally involving plastic deformation [3] under asperities covering a wide range of scales. The large plastic strains and strain gradients caused by the stress concentration in the specimen during sliding drive a highly localized dislocation activity and the formation of complicated dislocation patterns [4] near the surface. There have been attempts to adopt computational techniques involved with the characteristic length scale to quantitatively analyse the dependence of the frictional force on the contact size. Bhushan and Nosonovsky [5] adopted a strain gradient plasticity (SGP) model and demonstrated there are three zones of frictional stress dependence on the contact area. For small contact areas, frictional stress is found to be close to the shear strength of the specimen while for sufficient large contact areas, frictional stresses are determined by the Peierls stress. Between the two zones, frictional stresses are governed by the contact size and the plastic flow extending along the interface. On the other hand, featured with explicitly modelling the activities of individual dislocations, discrete dislocation plasticity (DDP) has been adopted to investigate the complex micro-sliding process. It was firstly used to study the micron-sliding size effect when contact size is too small to apply conventional plasticity [6]. Hurado and Kim [7] employed DDP analysis on the micro-sliding problem with the assumption that dislocations are only nucleated from the contact surface, and a three-regime variation of the shear stress based on the contact size similar to the one suggested in Ref. [5] was predicted. Deshpande et al. [8] applied two cohesive shear traction formulas on the contact in 2D-DDP simulations, and results show that both softening and non-softening cohesive constitutive equations generate similar frictional forces dependence trends, and the shear stress along a large contact size is in line with the dislocation source nucleation strength. The shear stress surge predicted at small contacts are validated in atomic force microscopic (AFM) tips [9]. The square root dependence of shear stress upon contact size has also observed in recent experimental measurements [10].

Turning now to more complex and computational demanding descriptions of material deformations, 3D-DDP [11] have also been used to describe the underlying mechanisms for both screw and edge types of dislocations transportation by virtue of indenter tip sliding. In the last decade there has been much activity related to the use of non-equilibrium Molecular Dynamics (MD) simulations, see *e.g.* [12], to address the origin of friction at the nano-scale and its links to what is perceived at larger time and length scales. Recent MD studies have revealed the deformation mechanisms responsible for surface and subsurface permanent deformation and microstructural changes in binary alloys, also exploring the temperature effect under dry sliding [13-16] at the atomic scale. Though the MD simulation results would be able to provide the meso-scale simulation with more fundamental information (e.g. stacking Fault energy that affects dislocation motion), the practical complexity of contact problems, such as the need to capture shear rate, details of the microstructure and surface roughness severely impedes MD simulations from accurately predicting the macroscopic performance of materials.

At the opposite end of the spectrum, crystal plasticity (CP) simulations address the deformation of crystalline materials under contact at the grain scale using a continuum approach and numerical (most often finite element) implementations. The time-efficient constitutive laws used in CP simulations enables to address the crystallographic deformation and damage mechanisms at the practical engineering scale, *e.g.* for a rolling contact fatigue [17] and galling [18] scenarios. However, the lack of length scale and resolution at the slip planes confines the CP modelling from further investigations into the physics-based mechanisms underpinned for microstructure change under the contact. We believe that a comprehensive examination of the microstructural changes by using discrete dislocation plasticity can help to further bridge the gap between the atomistic and macroscopic scales.

This work is further motivated by the evidence that ductile crystalline materials often exhibit deformed layers (termed as 'tribo-layers' or 'third body'[19]) arising from localized plasticity when subjecting to contact and tribological loads [20-24]. Simulations have been performed to understand the interaction between the contact and microstructural changes [13, 25-27]. However, no satisfactory interpretation of some of the emerging experimental evidence has been achieved so far; this is mainly linked to the complexity of the mechanisms regulating contact interactions and their intrinsically localised nature, which means multiple scales are involved and a single simulation technique cannot capture the materials behaviour. Recent experiments on face-canter-cubic (FCC) samples have observed an abrupt and highly localized microstructure change (termed as 'dislocation traceline' – DTL) [28, 29] at ~100nm depth from the contact during one-stroke sliding. This microstructural change eventually leads to recrystallization of the single crystal in subsequent cyclic loadings or when the material is more heavily loaded [30]. Approaches based on continuum assumptions [31] to identify microstructural changes, albeit providing good qualitative insight as to where such activity may take place, is hardly able to fully explain the mechanistic drivers for the formation of the dislocation traceline and the evolution of the discrete features associated with microstructural evolution. Obtaining mechanistic understanding at the dislocation length scale to capture the mechanisms responsible for surface and subsurface material evolution under micro-sliding is key to enable the prediction of the perceived macroscopic response of the material in terms of friction and wear evolution. Providing insight and tools to be incorporated in the design of new alloys will help tailoring material properties combining excellent friction and wear-resistance.

Although our previous integrated experimental and numerical investigations [30, 32] have unravelled the mechanisms driving the formation of the observed dislocation tracelines and microstructural changes, no comprehensive results and understanding has yet been provided to map the microstructural evolution of materials both at the surface and subsurface at the dislocation scale under dry sliding contact. In this paper, we examine the detailed plastic deformation at various depths to produce maps describing the likelihood and extent of permanent deformation and microstructural changes to occur at sliding interfaces using DDP simulation results. First, a comprehensive set of DDP simulations is conducted to understand the contact size effect on slip initiation with preceding indentation. The dislocation traceline and other microstructural changes in the subsurface under sliding are then shown to be associated to dislocations piling up and lattice rotation near the contact interface, which reflects recent experimental observations. In addition, the recrystallization observed in the sliding test under cyclic loading is shown to be driven by the development of geometrically necessary dislocation density and the resulting plastic strain energy density. Maps are provided to describe both the general behaviour of dislocation activities and the onset of development of discontinuities induced by contact to include a size effect.

# 2 Methodology
## 2.1 Discrete dislocation plasticity formulations

A planar, isotropic, isothermal discrete dislocation plasticity formulation firstly proposed by Van der Giessen and Needleman[33] is adapted here. The DDP formulations are fully addressed in earlier articles, *e.g.* [34, 35], hence only key points are concisely summarised here.

An FCC crystal structure is applied to specimens, with the plane of simulation taken perpendicular to crystal direction $[\bar{1}0\bar{1}]$ to satisfy the plane strain constraint. The material is assumed to be initially dislocation-free, and edge dislocations nucleate from Frank-Read sources, which are randomly distributed on the slip planes with a predefined density in the specimen. Dislocation activities within crystals are governed by constitutive laws, including mobility, pinning and escape from obstacles, which can be referred to [36]. Boundary conditions are satisfied using the superposition scheme first established in [37]. The fields of displacement, stress and strain are decomposed into a dislocation filed in an infinite elastic medium and a correction field that ensures the boundary conditions are satisfied; the former is obtained via summing up of analytical fields contributed from individual dislocations and the latter is obtained via a finite element solution of a boundary value problem where singularities are absent and thus dislocations' effect is mediated by the corrected boundary conditions. In this research, there are two types of numerical simulations using discrete dislocation plasticity, namely: sinusoidal indentation calculations and sliding calculations.

## 2.2 Sinusoidal indentation setup

Micro-indentation calculations are conducted on a film with thickness *H*=10μm (see Figure 1(a)) under a single sinusoidal shaped asperity. The asperity shape is characterized with the wavelength λ=10μm and the amplitude Δ=0.5μm. Dislocation activity is confined to a process window of dimension *l×H*=50μm×10μm. The process window is bounded at both left and right sides to an elastic region. The total width of the film is chosen sufficiently large as L=1000μm to avoid a boundary effect (*i.e.* trace surface condition at *x=±L/2*). Inspired by [38], three slip systems with $\Phi^{(\alpha)}$=0, ±45° with respect to *x*-axis, respectively, are assigned within the DDP process window. Aluminium-like material properties are assigned to the specimen, whose parameters are identical as and referred to [34]. The dislocation source density that indicates the different initial status of materials (*e.g.* pre-strain [39], heat treatment and pre-cracked, etc.) is fixed as $\rho_{nuc}$=48.5μm$^{-2}$ (the sensitivity study can be seen in Appendix) to minimize the dislocation source starvation effect [40]. The process window is discretised by a highly focused finite element towards the indenter lowest point. The finite element mesh is usually made up of 180×100 elements with a typical mesh size of 0.01μm in a sensitive zone of dimension 1μm×1μm. A time increment Δt=0.5ns is adopted to capture the dislocation activities.

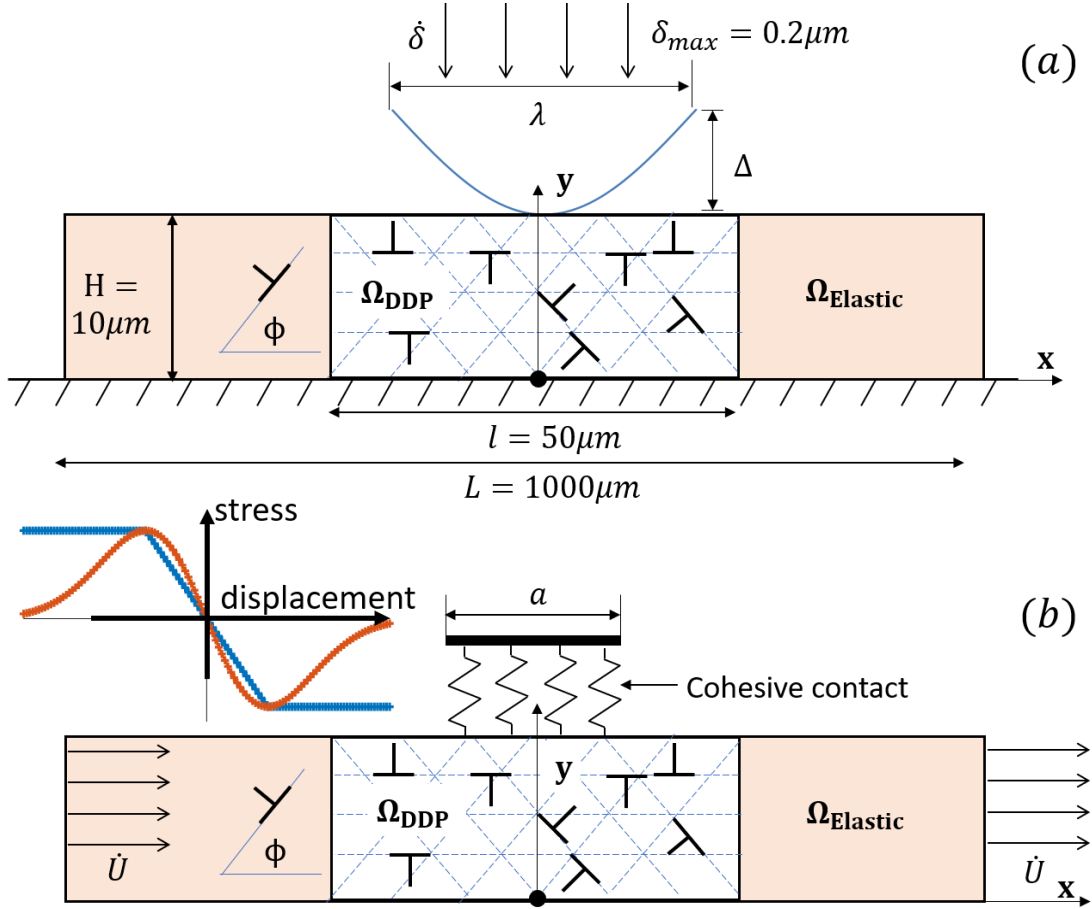

Figure 1. schematic diagrams of the (a) sinusoidal indentation and (b) sliding boundary value problems analyzed using discrete dislocation plasticity. The origin of the coordinate system employed is marked as a filled-in circle (•) in (a) and (b). The softening and non-softening cohesive relation between shear traction and jump displacement on contact partition is illustrated in (b). The non-softening relation is utilized in sliding calculations.

In sinusoidal indentation calculations, effects of geometry changes on the momentum balance and lattice rotations are neglected. However, the contact between indenter lower surface and film top surface is established on the deformed film surface. At an instant of indentation process, the instantaneous applied indentation depth $\delta$ is imposed on the rigid-body indenter. The corresponding actual contact length $A$ is defined as the range between the most left and right values of $x$ coordinates where the indenter contacts the deformed top surface. In general, the actual contact length $A$ differs from the nominal contact length $A_N = 2\lambda \cos^{-1}(1 - \delta/\Delta)$ due to sink-in or pile-up (see Ref. [41]), but it does not account for surface roughness (as analysed and discussed in [42]), which could lead to a significantly smaller contact area and hence spikes in indentation pressure due to random fluctuations along the contact, especially when sharp indenters are involved. The maximum indentation depth in this research is limited as $\delta_{max}$=0.2µm which is sufficiently small compared to the film thickness (relative indentation depth 0.02) to preventing the rigid substrate from taking its effect to disturb the film response [43] during the micro-indentation process.

The boundary conditions of sinusoidal dentation problem analysed using DDP formulations are detailed as following:

$$\dot{u}_1 = 0, \dot{u}_2 = \dot{\delta} \text{ on } S_{contact}$$
$$\dot{u}_1 = 0 \text{ on } y = 0 \text{ and } \dot{u}_2 = 0 \text{ on } x = 0 \qquad (1)$$
$$T_1 = T_2 = 0 \text{ on } y = H \notin S_{contact}$$

Where $u_i$ is the displacement component, $S_{contact}$ the contacted fraction of the top surface and $T_i = \sigma_{ij}n_j$ the surface traction on a surface with normal vector $n_j$. The displacement rate of the indenter is set as $\dot{u}_2 = \dot{\delta} = 0.4 \text{ ms}^{-1}$.

The total reaction force of film response to the indenter penetration is computed as:

$$F = -\int_{-A/2}^{A/2} T_2(x, H) dx \qquad (2)$$

The actual indentation pressure $p_A$ is defined by:

$$p_A = F/A \qquad (3)$$

## 2.3 Sliding simulation setup

The specimen dimension and the material properties used in the following sliding calculations are identical as those in the indentation (see Figure 1(b)). The contact between the sinusoidal asperity and specimen is modelled via a resistant adhesion zone on the contacting surface of actual length $A$ with a relation between shear traction versus displacement, which is given by:

$$T_t = \begin{cases} -\tau_{max}\frac{\Delta t}{\delta_t}, if \ |\Delta t| < \delta_t \\ -\tau_{max} sign(\Delta t), if \ |\Delta t| > \delta_t \end{cases} \qquad (4)$$

Where $\Delta t = u_1(x, H)$ is the tangential displacement jump across the cohesive surface, $\delta_t$ the critical jump, $\tau_{max}$ the cohesive strength, and $T_t$ represents the shear traction response.

The maximum cohesive strength $\tau_{max}$ is as τmax=300MPa and the threshold displacement jump is δt=0.5nm. Those values are identical as in previous work in [34] to understand the dependence of shear stress upon contact size at different regimes. Regarding the cohesive relation between the shear traction and the displacement jump, Deshpande et al [8] applied another form of cohesive equation to represent a softening cohesive relation:

$$T_t = -\sqrt{e}\tau_{max}\frac{\Delta t}{\delta_t}\exp\left(-\frac{\Delta t^2}{2\delta_t^2}\right) \quad (5)$$

The softening and non-softening cohesive shear displacement relations are compared in Figure 1(b). They concluded that the onset of sliding along contact does not strongly depend on the form of the cohesive relation. Therefore, we employ the non-softening form in this research to help convergence. Also the temperature change due to friction and its influence on dislocation activities [39, 44] is neglected in the research.

The sliding rates,

$$\dot{U}_x = \dot{U}, \dot{U}_y = 0 \quad (6)$$

are imposed on the specimen boundaries $x=\pm L/2$ and $y=0$ to simulate the relative displacement of contact surface with the rate $\dot{U}/A = 10^4 \text{s}^{-1}$, which substantially rules out the sliding rate sensitivity [45] that is shown in Appendix (see Figure A1).

The averaged shear stress $\tau$ along the contact is given by:

$$\tau = -\frac{1}{A}\int_{-A/2}^{A/2} T_x(x,H)dx \quad (7)$$

In one set of sliding calculations, the film subjects to a pure shear slide condition with an initially indentation depth free (and hence dislocation and stress free) status. Shear stress along with predefined contact can be studied without normal stress as the work in [8, 34, 45]. However, this set of calculations can only approximate the sliding scenario as the local material deformation near the contact surface and subsurface by virtue of normal load that has not yet been taken into consideration. Issues including surface elevated [34] during sliding given rise to the absence of normal stress also limits the application of pure shear sliding. In the other set of sliding calculations, a sinusoidal indentation simulation is firstly carried out to establish contact, dislocation structure and initial fields with a certain indentation depth for forthcoming sliding calculations. The latter set of simulations aims to capture more realistic features of material deformation under sliding with a certain normal load and initial dislocation structures, the latter of which has been studied in the micro-indentation problem [43]. Results obtained from two sets of sliding simulations are compared to reveal the effect of preceding indentation load.

## 3 Sinusoidal indentation

We start by exploring features associated with the indentation pressure and contact size variation of the indentation response originally reported in Ref.[32]. The normal stress field with corresponding dislocation structure for four indentation depths during the sinusoidal indentation process described in Section 2 are detailed in Figure 2.

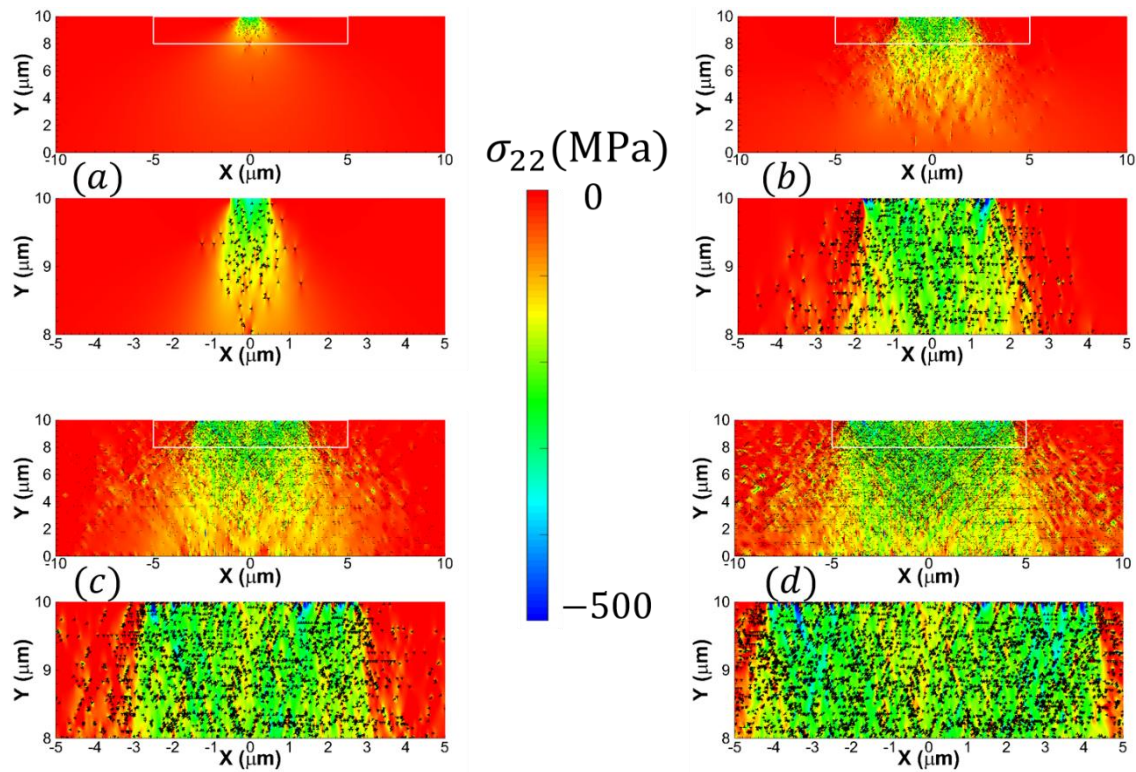

Figure 2. Normal stress σ$_{22}$ and the corresponding dislocation structure at indentation depths (a) δ=0.01μm (b) δ=0.05μm (c) δ=0.10μm, and (d) δ=0.20μm, respectively, for a sinusoidal asperity with λ=10μm and Δ=0.5μm. For each indentation depth, the upper subfigure shows a full view of and the lower subfigure shows the subsurface region denoted by a white frame in the full view subfigure.

For each indentation depth, a full view of the stress distribution within the film and a magnified view of the region near the indenter are reported in the upper and lower subfigures, respectively. Plastic flow expands from the surface into the bulk with an increase in applied indentation depth. The normal stress generally exhibits a more homogenous distribution underneath the contact, which differs from that obtained in typical wedge-shaped indentation [46]. Hence, the indentation size effect, observed as an increase in the indentation pressure at sufficiently small indentation depths, is diminished in sinusoidal indentation by virtue of the relative absence of a strain (stress) gradient under the contact compared to indentation by a wedge. The initial stress fields, dislocation structure and contact sizes ($A$=1.11, 3.55, 5.52, 8.42μm) caused by the indentations shown in Figure 2 provide the starting points for the following sliding simulations.

## 4 Contact size effect on slip initiation and full slip

The average shear stress $\tau_{avg}$ developed during the sliding along the contact surface is shown in Figure 3 for the four contact sizes achieved by the sinusoidal indentation depths described in Section 3. The sliding simulations were performed sufficiently slow to eliminate the sliding rate sensitivity that was discussed in Ref. [45].

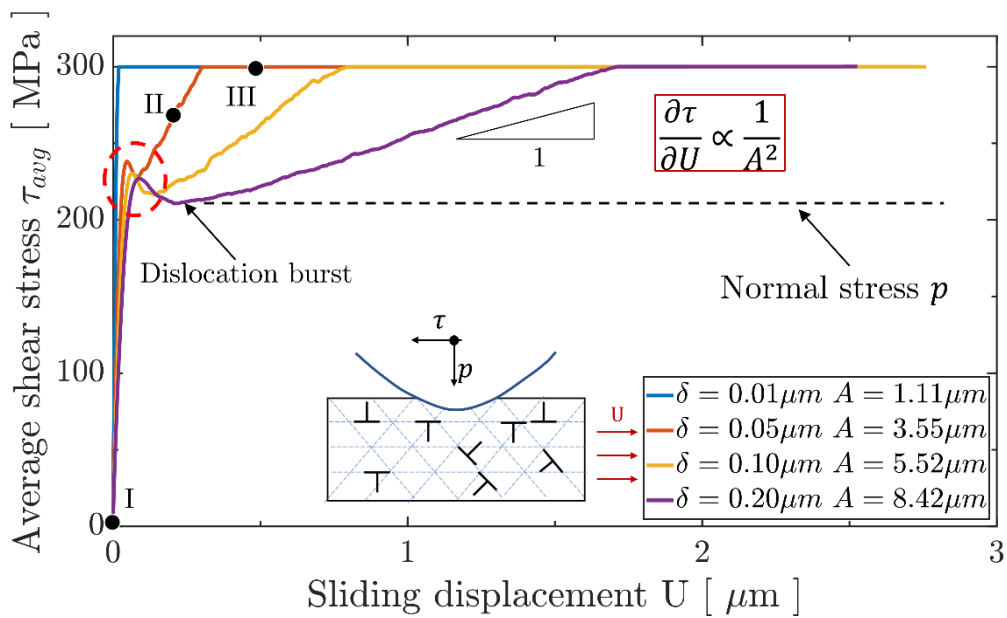

Figure 3. Shear stress averaged along the contact, $\tau_{avg}$, versus sliding displacement $U$ for different contact sizes introduced by preceding sinusoidal indentation. The normal stress $p$ is denoted by a black dashed line.

The average shear stress increases elastically with sliding distance $U$, which is followed by a temporary oscillation (denoted by the red dashed circle in Figure 3) that is produced by dislocation nucleation bursts and associated stress drops [47]. The shear stress continues to increase thereafter, but at a reduced rate (e.g. see location denoted by II on Figure 3) due to plasticity arising from dislocation activity, except for the smallest contact size case where an insufficient number of dislocations are nucleated (see Figure 2(a)). The shear stress eventually plateaus at the defined cohesive stress of the contact (*i.e.* 300 MPa for the set of simulations presented here). The slope of the curve in the hardening region (II) is found to be inversely proportional to the square of the contact size, with larger contact sizes requiring a longer sliding distance to achieve slip at the contact surface (*i.e.* when the shear stress reaches the cohesive strength). The shear stress distribution caused by dislocation activity underneath the contact is detailed for representative indentation depths and the different sliding regimes (denoted by the points I, II and III) in the following sections.

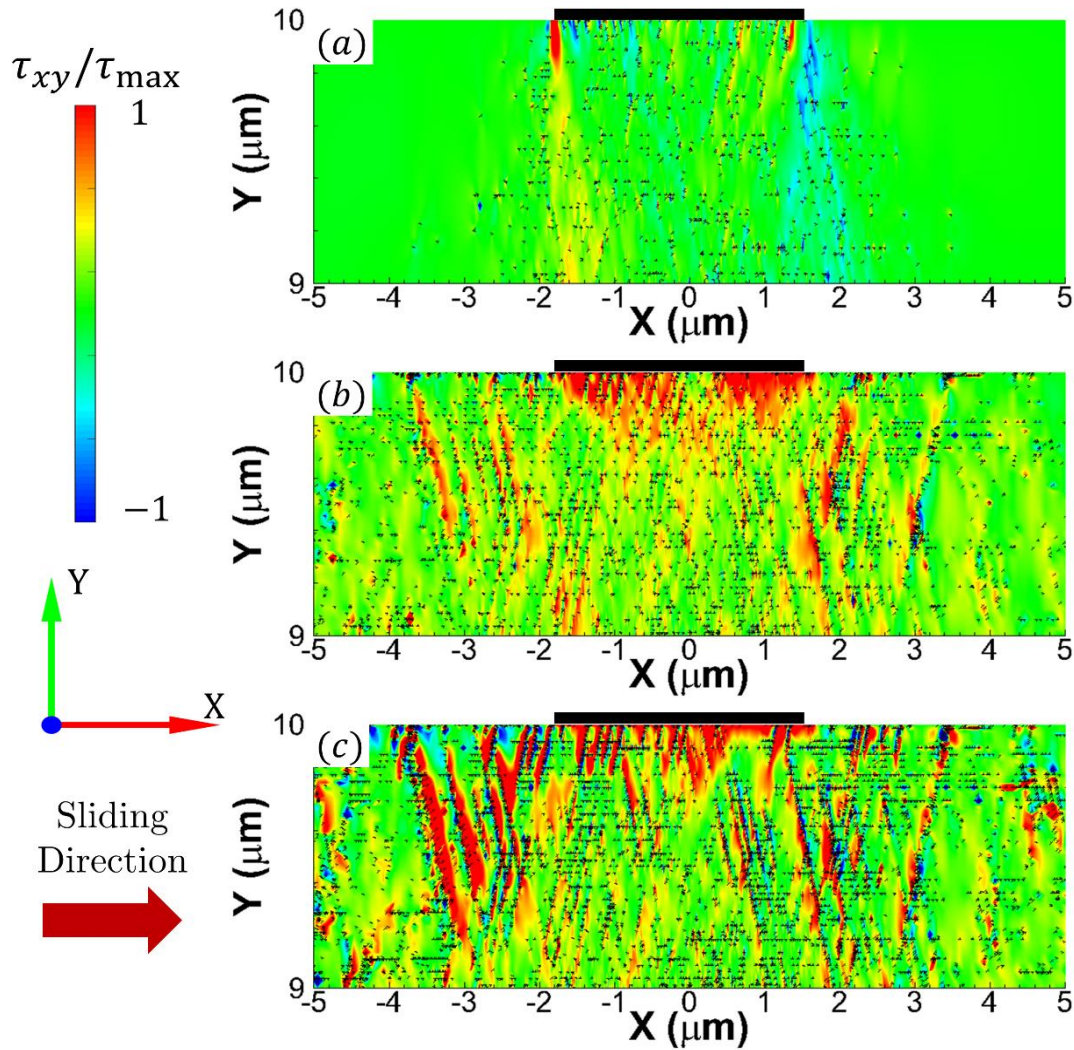

Figure 4. Normalized (by the maximum cohesive strength) shear stress distribution with the corresponding dislocation structure for contact size $A$=3.55µm, introduced by preceding sinusoidal indentation to $\delta$=0.05µm, at instants when the sliding distance equals: (a) $U$=0µm *i.e.* immediately after the indentation, (b) $U$=0.18µm and (c) $U$=0.54µm. These three instants are representative states for different stages of sliding initiation at the surface (i.e. corresponding to regions I, II and III in Figure 3).

The shear stress distribution with instantaneous dislocation structure for contact size *A*=3.55µm is shown in Figure 4 for three different sliding distances. The shear stress is highly localized at the contact edges only when the sliding process starts at *U*=0 (point I in Figure 3) in Figure 4(a). The shear stress develops from the edges towards the centre of the contact while dislocations are nucleated and glide into the bulk of the specimen at *U*=0.18µm in Figure 4(b), which corresponds to the secondary ramp (point II in Figure 3) in the average shear stress evolution; slip occurs on the corresponding part of the contact surface.

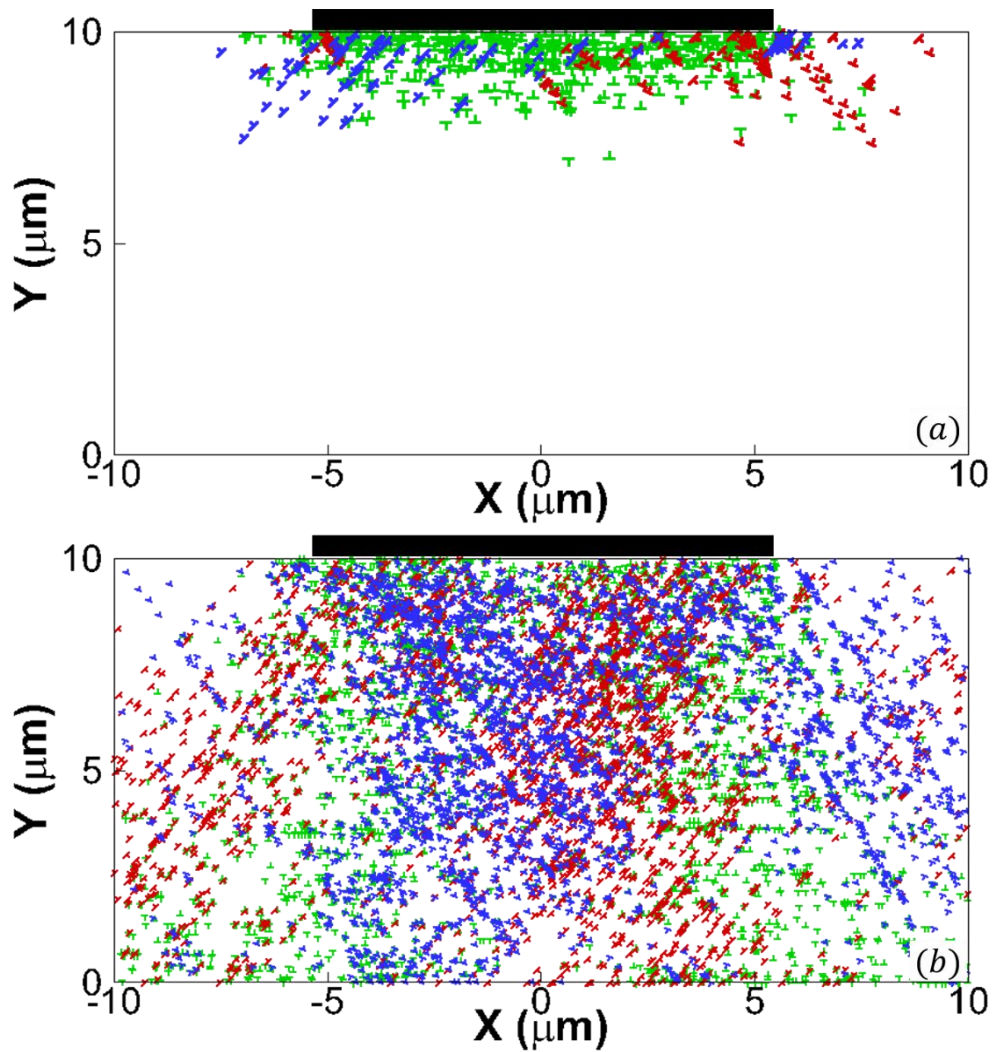

Figure 5. Instantaneous dislocation structure at the instant of sliding distance $U \cong 0.10$µm (a) sliding without prior indentation and (b) sliding with prior sinusoidal indentation. Contact sizes are selected to be identical as *U*=9.8µm for both cases. Dislocation symbols denoted by different colours stand for dislocations gliding along three different slip systems.

With further sliding, the shear stress fully saturates at the maximum cohesive strength over the entire contact region when *U*=0.54 (point III in Figure 3), as shown in Figure 4(c); hence, the average shear stress does not evolve with the sliding distance beyond the final transition to full sliding. The whole contact surface initiates slip at this point in the sliding, as the cohesive strength can no longer sustain the shear stress.

The average shear stress is shown (see Figure A2) to be independent of the dislocation source strength, even for the larger contact size, which was first revealed in other contact size studies of a pure sliding configuration (*i.e.* without preceding indentation), see e.g. Refs. [8, 34]. The dislocation structures at the same sliding distance and similar contact size are compared in Figure 5 for the pure sliding and sliding with preceding indentation cases. For the pure sliding case (Figure 5(a)), dislocations accumulate only near to the contact surface; the region far away from the contact is free of dislocations. However, dislocation activity occurs everywhere when indentation precedes sliding (Figure 5(b)). The dislocation density in the specimen that experiences indentation before sliding is $\rho_{dis}=21.2\mu m^{-2}$, which is 20 times higher than that without indentation, $\rho_{dis}=1.40\mu m^{-2}$. The widespread plasticity introduced by the preceding indentation significantly changes the material response, hence its resistance to sliding. Therefore, the dependence of the shear stress upon contact size not only arises from the plasticity due to the contact itself, but also from the contribution of the prior indentation; the latter provides the initial dislocation structure, which in turn may significantly affect the sliding process.

The total and geometrically necessary dislocation (GND) density (calculated via the open Burger's circuit method [48, 49]) evolution during sliding is shown in Figure 6(a) and (b) for the four contact sizes established by the prior indentation. The total dislocation density (Figure 6(a)) for all contact sizes increases linearly with sliding distance initially. The rate of increase then reduces, and eventually reaches a steady value (except for the largest contact size *A*=8.42um) at a critical sliding distance (indicated on the figure by dashed vertical lines) that corresponds to the subsurface plastic flow relative to the contact. The critical sliding distance shows a strong positive dependence on the contact size, which is rationalized by the size effect of the plasticity underneath the contact [10]. To exclude the effect of the dislocation density introduced by the indentation [46, 50], the fraction of the dislocation density that is attributable to GNDs is examined in Figure 6(b), which again shows a strong positive dependence on contact size and achieves a steady rate of increase at the critical sliding distance.

The dependence of the average shear stress and the dislocation density on the contact size suggests that a map identifying the conditions under which slip initiates and full sliding is achieved can be determined using the simulation results, which can then be qualitatively applied to a number of crystalline systems. The material response is governed by both the preceding indentation (in terms of indentation depth $\delta$) and the sliding (in terms of sliding distance $U$). The four representative indentation depths (*i.e.* various contact sizes) that are discussed above are labelled as black dashed arrows in Figure 7. Different zones are categorized by the onset of surface slip (followed by partial slip), full surface slip and subsurface plastic flow, which are obtained by the system response in terms of average shear stress and dislocation density. For a given indentation depth (*i.e.* a fixed contact size), slip initiates at the contact edges (Zone A), gradually spreads to the whole contact (Zone B), completely occupies the entire contact surface (Zone C) and spreads into the specimen bulk (Zone D). These zones could also be linked to contact adhesion, and its interplay with tangential stresses in the presence of material non-linearities [51-53], and the evolution of contact partial slip and slip zones, which play a significant role in controlling the wear behaviour of alloy surface. Although a direct link between wear and material deformation cannot be easily established, the proposed maps can also be used to assess the likelihood of occurrence and severity of wear[13, 15, 54].

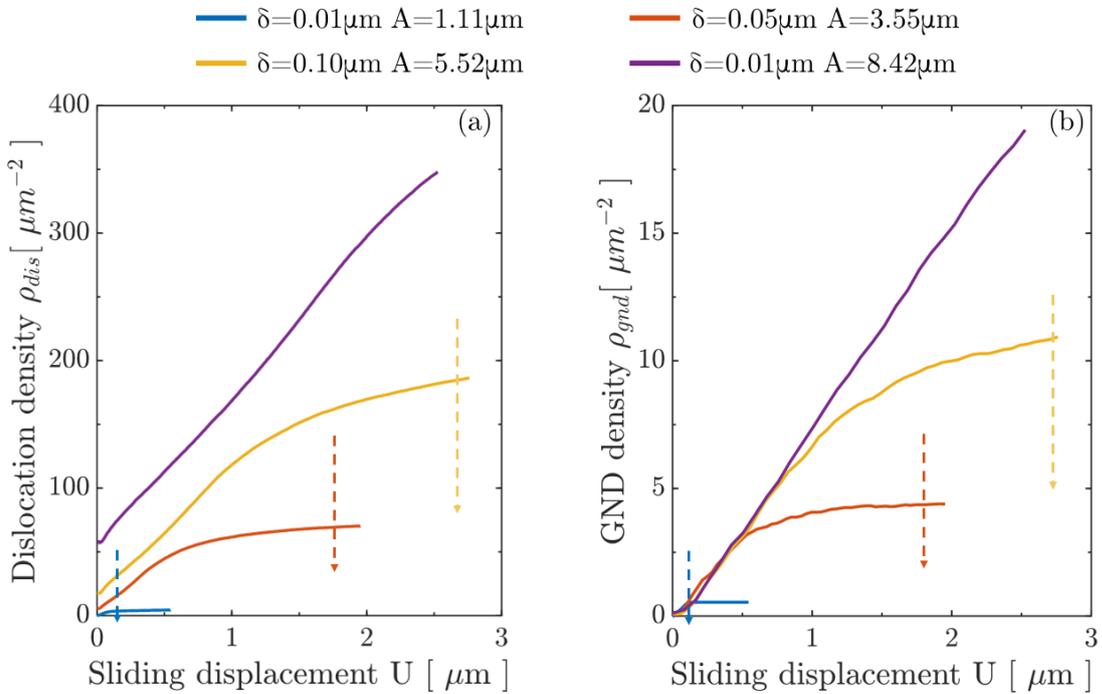

Figure 6. Evolution of density of (a) total dislocations $\rho_{dis}$ and (b) geometrically necessary dislocations $\rho_{GND}$ with sliding displacement $U$ under different contact sizes.

For a given sliding distance, a larger contact size substantially delays the onset and fulfilment of slip between the indenter and the specimen, which is in line with other experimental observation [55] and numerical predictions [56]. This should in turn increase the lifetime of materials under fretting fatigue [57].

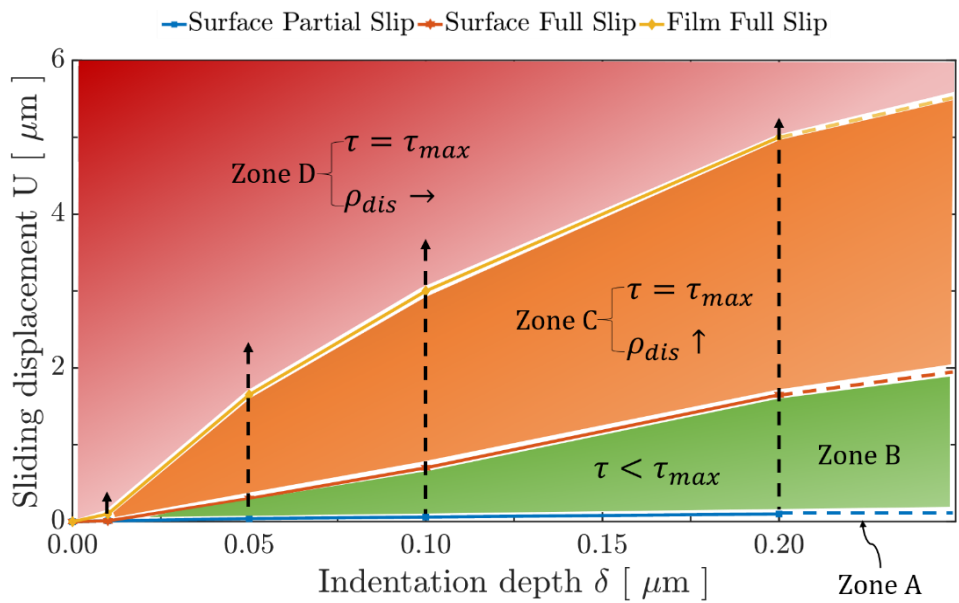

Figure 7. Sliding distance against indentation depth map demarcated into: white regions (Zone A, no slip), green region (Zone B, contact surface partial slip), orange region (Zone C, contact surface full slip) and red region (Zone D, subsurface plastic flow, loss of contact constraint).

## 5  Contact size effect on microstructural changes

The shear stress field at the instant when full slip is achieved for the specimen, or the maximum sliding distance is achieved (for the largest contact size $A$=8.42µm only), is shown with the deformed material configuration in Figure 8 for the four contact sizes. The surface is significantly deformed at the leading edge of the contact, where the height of material pile-up that originates from dislocation activity in the subsurface is estimated to be around 0.2µm in the case of $A$=5.52µm and 8.42µm. Negative shear stress that comes from the material's resistance to the pile-up is also observed, which may potentially serve as a crack initiation precursor, along with the homogenous deformation that occurs during excessive contact sliding or cyclic sliding.

The 'plough effect' [3] that occurs when the contact moves along the surface introduces additional roughness to the contact [58] and eventually leads to local damage initiation and short crack nucleation (at the scale of individual crystals) or even macroscopic failure. In addition, a large contact size also introduces a deep and wide plastic deformation zone underneath the contact and into the bulk of the material, which may lead to subsurface material damage and microstructure transformations during sliding.

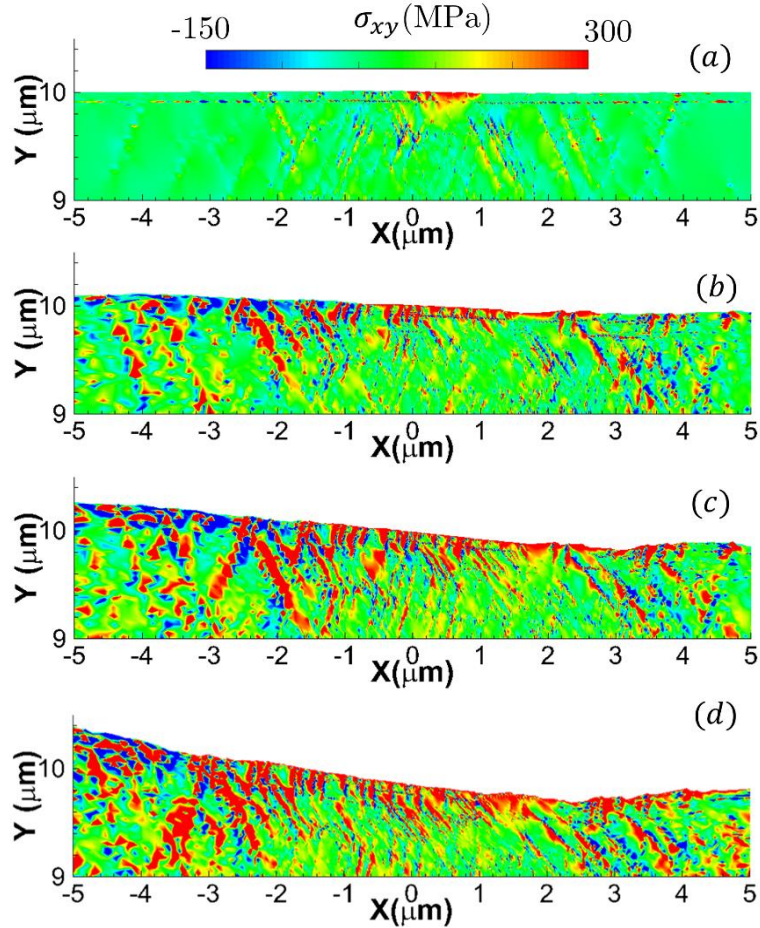

Figure 8. Shear stress field in the film and the deformed surface profile (with no displacement magnification) for contact sizes (a) *A*=1.11μm, (b) *A*=3.55μm, (c) *A*=5.52μm and (d) *A*=8.42μm. Results are shown at the instant when full slip or maximum sliding distance has been achieved.

The surface and subsurface damage can also be evaluated by closely looking at the amount of slip generated by dislocation motion that is driven by the sliding load. Independent experimental observations have shown that slip is the key mechanistic precursor for fatigue crack nucleation in FCC crystalline materials [59, 60]. The total slip is calculated as $\Gamma = \sum_{\alpha=1}^{3} \left| s_i^{(\alpha)} \epsilon_{ij} n_j^{(\alpha)} \right|$, where ***ε*** is the strain tensor and ***s***[(α)] and ***n***[(α)] are respectively the unit vectors in the directions of slip and the slip plane normal for the $\alpha^{th}$ slip system [36]. The total slip at the instants when full slip or maximum sliding distance is achieved are shown in Figure 9 for the four contact sizes: total slip is highly localized near the contact (especially at the leading edge) and a large contact size results in a large amount of plastic slip, both at the surface and in the bulk of the specimen, which may be an indicator for material failure under sliding conditions. The intense horizontal bands of slip evident in Figure 9 are consistent with recent high-resolution digital image correlation (HR-DIC) experimental observations (e.g. [61]); discrete features such as this are not predicted by crystal plasticity (e.g. [62]).

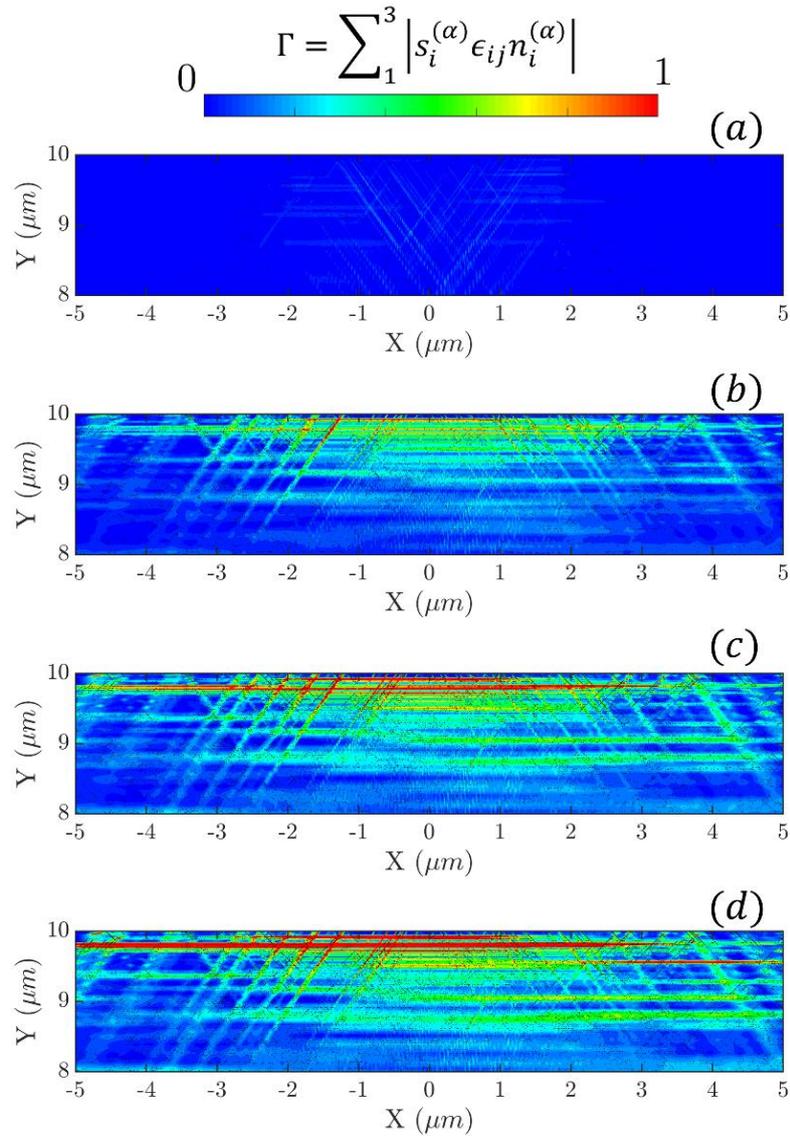

Figure 9. Total slip field in the undeformed film for contact sizes (a) *A*=1.11μm, (b) *A*=3.55μm, (c) *A*=5.52μm and (d) *A*=8.42μm. Results are shown at the instant when full slip or maximum sliding distance has been achieved.

Lattice rotation has been linked to the observed dislocation traceline [32] and subsequent microstructure change under dry sliding; lattice rotation bands were correlated with zones of incipient microstructure change. Three depths (all bands are approximately parallel to the sliding direction) are proposed to characterize the dimension of the zone of microstructure change, as shown in Figure 10(a) for contact size *A*=3.55μm. The first, $h_0$, denotes a zone of low lattice rotation extending from the contact surface into the bulk, which has been shown to be independent of contact size and consistent with experimental measurements [28]. The others, $h_1$ and $h_2$, bound a zone of high lattice rotation, underneath the low lattice rotation zone above it; the location and size of the high lattice rotation zone depends strongly on the contact size, which is created by the prior indentation.

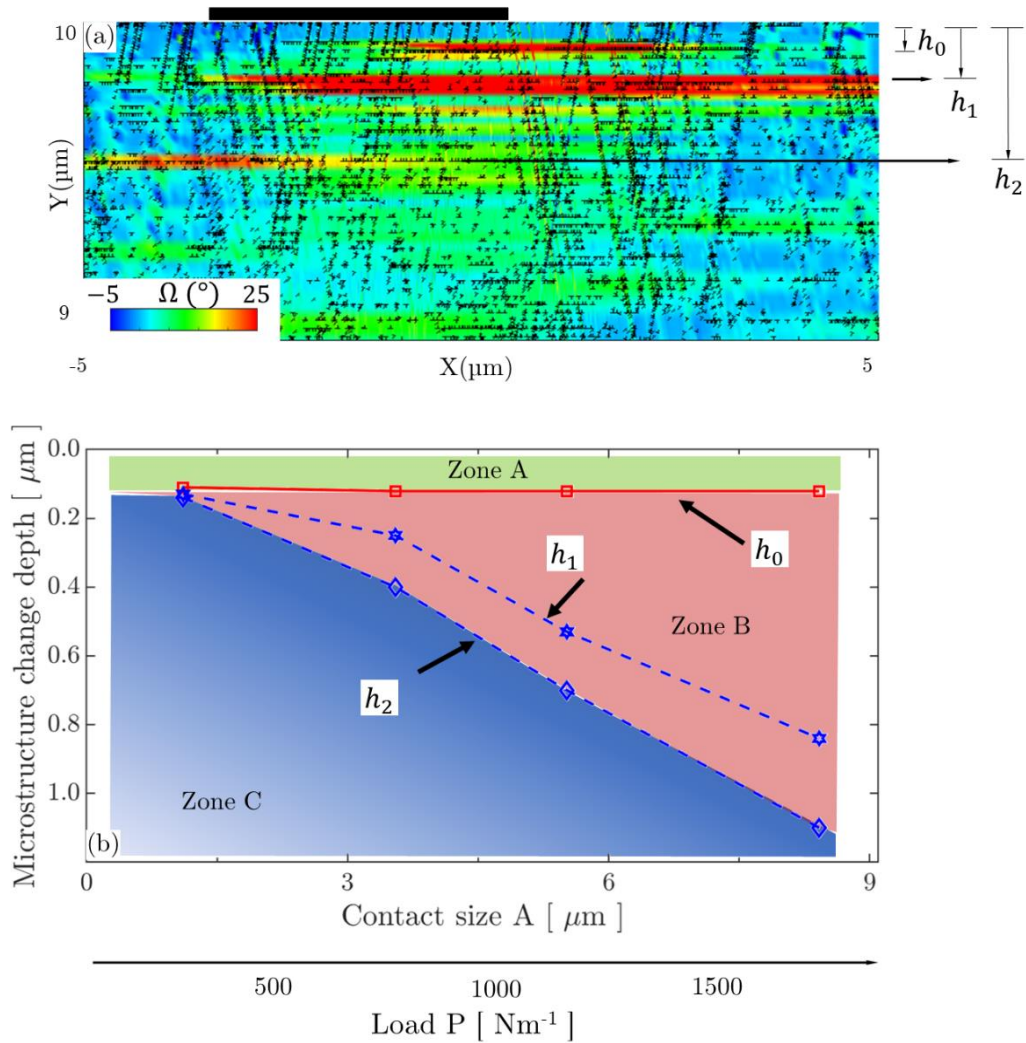

Figure 10. Lattice rotation indicators for microstructure change. (a) Dimensions that characterize bands of relatively uniform lattice rotation, shown for *A*=3.55μm. (b) Microstructure change map based on lattice rotation distribution. Zone A (green): surface region constrained by the contact with low lattice rotation and limited microstructure change predicted; Zone B (red): subsurface region with high lattice rotation, microstructure change predicted; Zone C (blue), bulk material with low lattice rotation, no microstructure change predicted.

A map, which provides an indication of the regions where microstructural changes are likely for a given contact size, can be constructed using these three characteristic depths and the four contact sizes used in the simulations (see Figure 10(b)). The three zones categorize the emergence of material layers under sliding contact in terms of the lattice rotations that they would be subjected to, with Zone B the most severely-rotated layer. It must be noted here that things may differ for scenarios in which relatively large loads are applied. In such circumstances, as reported in [63], multiple DTLs may form near the contact surface and larger rotations and formation of subgrains may take place. This is due to the increase of the energy made available to the material to trigger various microstructural transformations, also linked to the formation of twin boundaries [13]. This means that, in such scenarios dealing with larger loads, Zone A in Figure 10(b) may reduce or disappear as the deformation mechanisms describe in Zone B delocalise to reach much larger regions of the material and also migrate towards the surface. This map indicates the onset of microstructural change driven by plastic deformation at the dislocation scale of a crystalline material, as a function of the external tribological loading conditions. A map such as this could be used to optimise surface performance (*e.g.* wear resistance) by tailoring surface properties and initial microstructure in order to compensate for the effects of in-service microstructure change.

In addition to the lattice rotation shown above, geometrically necessary dislocation (GND) density can be used to identify the regions of potential microstructure change [64-66], as GNDs compensate for the lattice curvature during plastic deformation [67, 68]. The GND density distribution is calculated based on the net Burger's vector algorithm [69] for the four contact sizes. Larger contact sizes introduce a GND density concentration that extends deeper under the contact (see Figure A3 in the Appendix). The band of high GND density is characterized by the depths $h_0$ and $h_{max}$, as shown in Figure 11(a) for contact size $A$=3.55µm. The depth $h_0$ at which the band of high GND density begins is independent of the contact size, whereas the maximum depth $h_{max}$ shows a strong positive dependence on the contact size (i.e. normal load). A microstructure change map can also be constructed from the bounds on the high GND density band for the four contact sizes, which is depicted in Figure 11(b); it is consistent with the map shown in Figure 10(b).

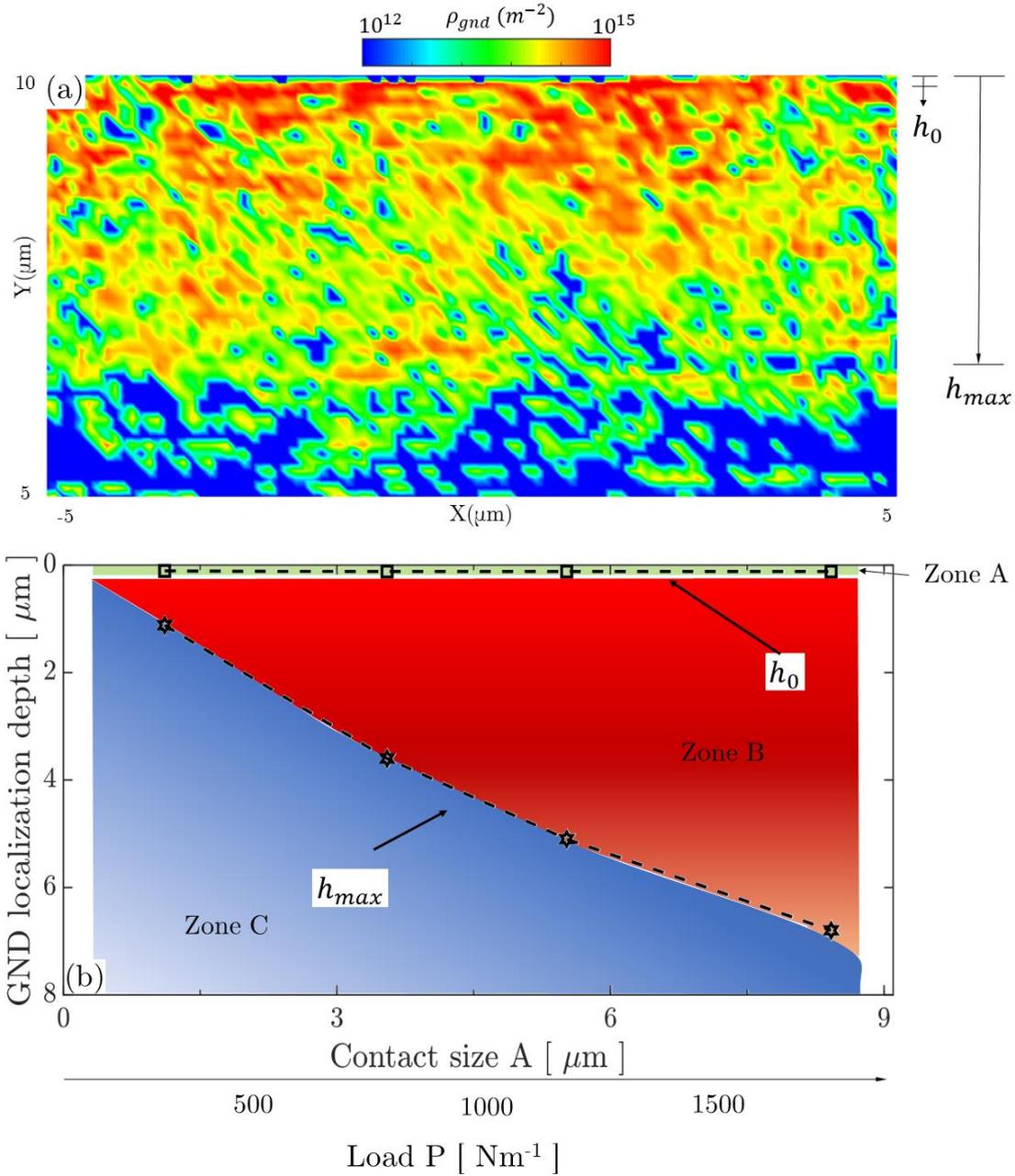

Figure 11. GND density indicators for microstructure change. (a) Dimensions that characterize the band of high GND density, shown for *A*=3.55µm. (b) Microstructure change map based on GND density distribution. Zone A (green): surface region constrained by the contact with low GND density, limited microstructure change predicted; Zone B (red): subsurface region with high plastic curvature hence GND density, microstructure change predicted; Zone C (blue), bulk material with low GND density, no microstructure change predicted.

Recrystallization, rather than crack nucleation, was observed in the specimen under sliding when a relatively large normal load was applied [30]. The stored energy associated with the dislocation structure, called plastic strain energy density (PSED) here, ($U = \boldsymbol{\sigma} : \tilde{\boldsymbol{\epsilon}}$, where σ and $\tilde{\epsilon}$ are the stress and plastic strain tensor, respectively) is calculated within the material under the four contact sizes investigated in this paper; the case with contact size A=3.55µm is shown in Figure 12(a) as an example. The layer immediately beneath the surface (indicated by the white arrows in Figure 12(a)) shows strong PSED concentration, which is produced by dislocation pile-ups in that region. In addition, the highest PSED is found to be in the subsurface rather than at the surface, which is a result of the locking and constraining effect of the contact [32].

The averaged PSED (the width used to compute the volumetric average is chosen as twice the contact size) plotted against the distance away from the contact (*i.e.* perpendicular to the sliding direction) is reported in Figure 12(b) for the four contact sizes (symbols). A Gaussian fit (solid lines) is used to represent the distribution of PSED as a function of depth under the contact. The PSDE increases to a peak value in the subsurface and then decrease to a plateau value in the material bulk for all contact sizes; as expected, small contact sizes show a much sharper peak, indicating greater localisation. The peak depth is found to be 100-200nm from the contact for all four contact sizes, which correlates well with the experimental observation on the microstructural changes reported in Ref. [28]; it should be noted that this is far away from the location of the maximum shear stress depth (~20µm) predicted using Hertzian theory[31]. With the increase of contact size, the PSED tends to develop to a greater extent in the subsurface, which corresponds to more energy available for *e.g.* microstructure change. A larger contact size also results in a smoother distribution with depth. Therefore, it is postulated here that an even larger contact size *e.g.* closer to the one applied in the experimental test (A≈90µm) reported in Ref. [28], would lead to a larger (in depth) high PSED zone. If may follow that the application of cyclic sliding, which periodically supplies energy and incrementally promotes plastic deformation, would eventually result in recrystallization, hence permanent microstructural change in the regions of the material under the contact corresponding to high PSED. Hence, the underpinning mechanism for the recrystallization observed in the experiments appears to be linked to the PSED resulting from the dislocation activity and structure in the material in the immediate proximity of the contacting surface. This implies that a larger load would lead to a larger region of microstructure change, down into the bulk of the material, as recently observed by Molecular Dynamics [13, 16]. Hence the PSED profile indicates the potential recrystallization zone for various contact sizes, although detailed quantification of the extent of re-crystallisation cannot be captured by the DDP analysis in isolation.

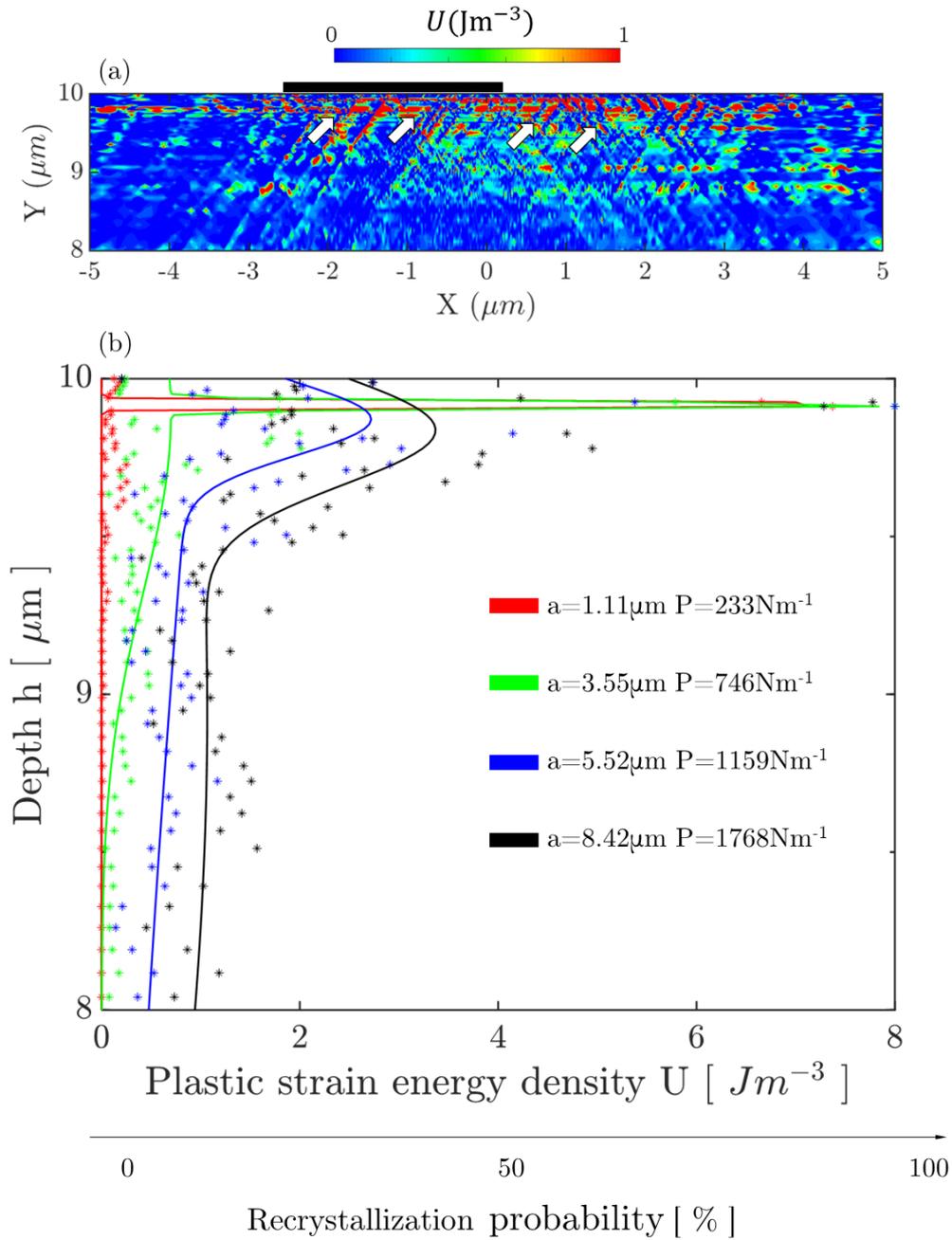

Figure 12. (a) The plastic strain energy density (PSDE) localization zone under the contact size A=3.55μm, where full slip is achieved. (b) The profile of average PSDE through the depth of material under the four contact sizes (normal load).

# 6  Conclusion

Nano- and micro-sliding analyses were carried out at a single asperity scale, where discrete dislocation plasticity is used to extensively investigate the emergence of permanent deformation and associated plastic strain energy stored in the dislocation structure within a single crystal subjected to contact and frictional sliding. This investigation explores surface slip as well as subsurface plastic flow, crystallographic slip, dislocation activities and their interplay. The following conclusive points have been drawn:

- The surface slip and subsurface plastic flow are found to be strongly dependent on the contact size and the sliding distance, and a map that depicts the correlation between surface slip and contact size/indentation depth has been established (Figure 7).
- The mechanisms for subsurface microstructural evolution that was observed in independent experiments has been associated to microstructural changes induced by localized lattice rotation and plastic strain energy, which results from dislocation piling up. The extension and localisation of the plastic strain energy density is strongly affected by the contact conditions and applied load.
- Two maps that describe the contact-size dependent microstructural evolution (Figures 10 and 11), which is governed by the dislocation density, have been proposed. They facilitate the optimization of material response and surface performance under sliding.

# Acknowledgement

This work was supported by the Engineering and Physical Sciences Research Council (EPSRC) [Reference: EP/N025954/1].

# Appendix

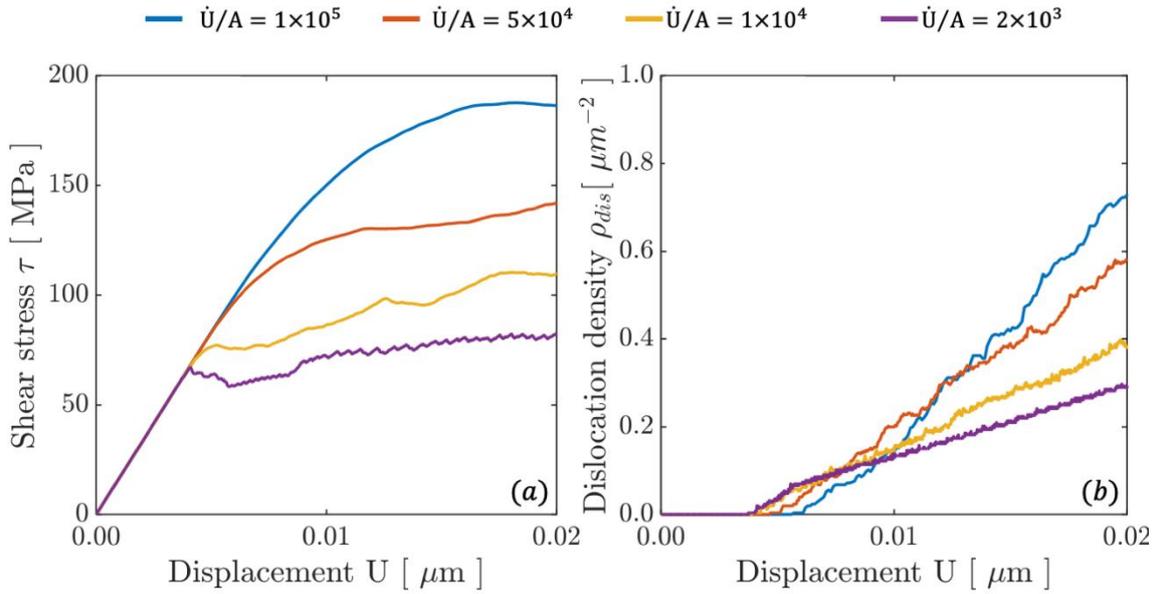

Figure A1. (a) the shear stress and (b) dislocation density against sliding distance under various sliding rates. The sensitivity study was conducted with the contact size 0.6μm and without preceding indentation.

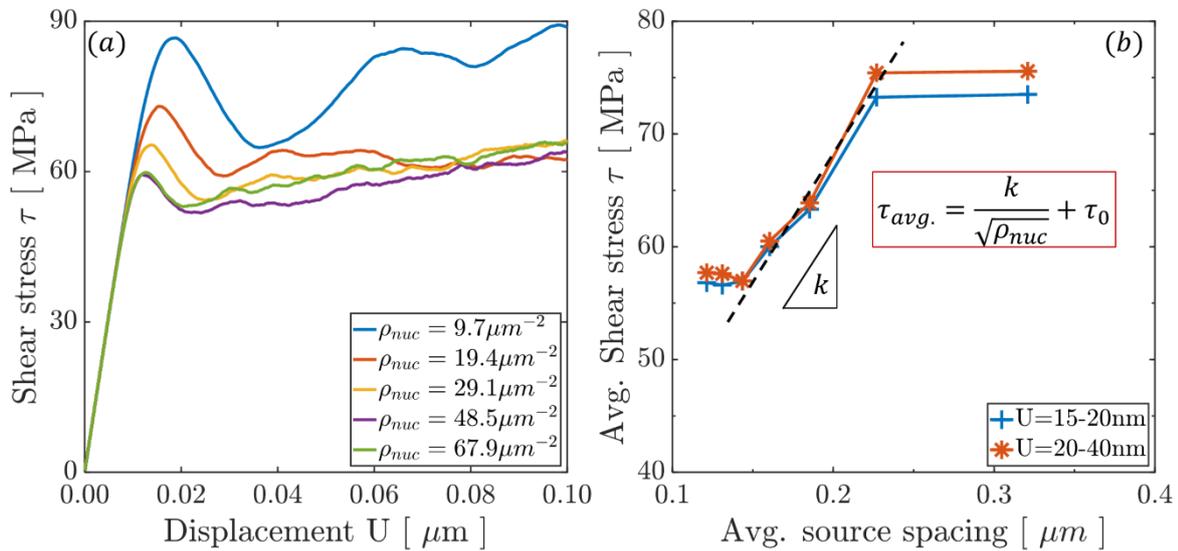

Figure A2. (a) shear stress τ versus sliding displacement U curves for selected values of dislocation source density $\rho_{nuc}$ with contact size A=4.0μm. (b) the relation between average stress stress $\tau_{ave}$ amd average source spacing. Results are obtained from sliding simualtions with no prior indentation.

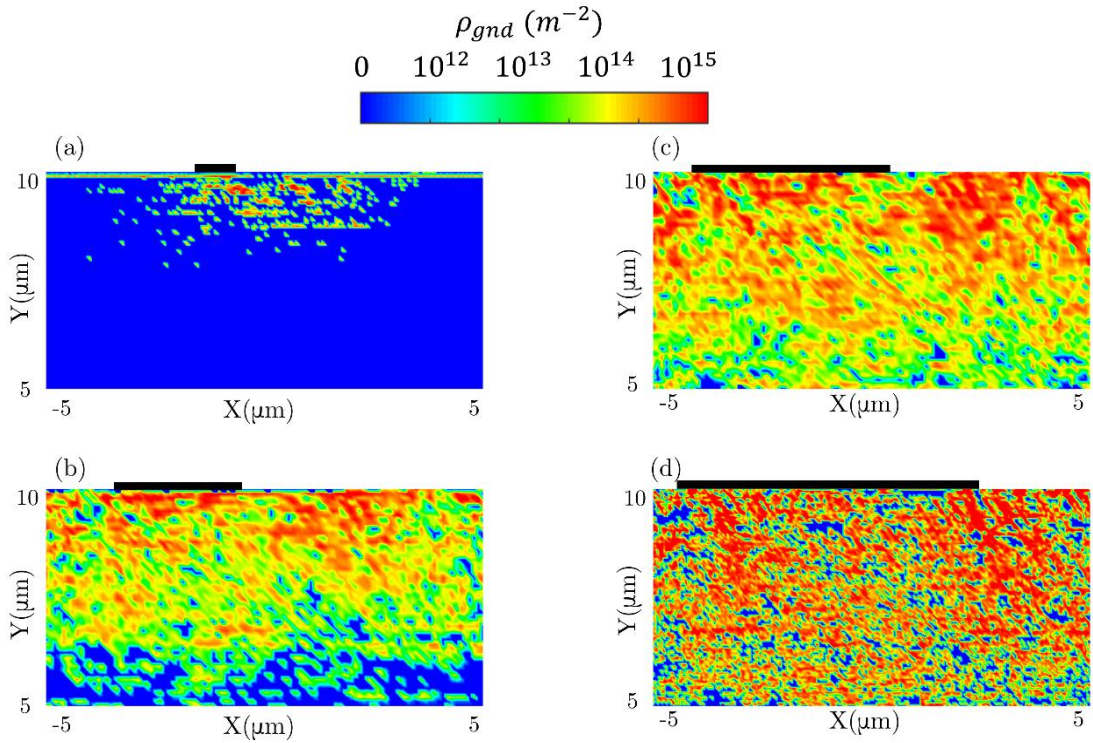

Figure A3. geometrical necessary dislocation (GND) density field in the undeformed film under contact size (a) A=1.11μm (b) A=3.55μm (c) A=5.52μm (d) A=8.42μm. Results are shown when full slip or maximum sliding distance has been achieved.

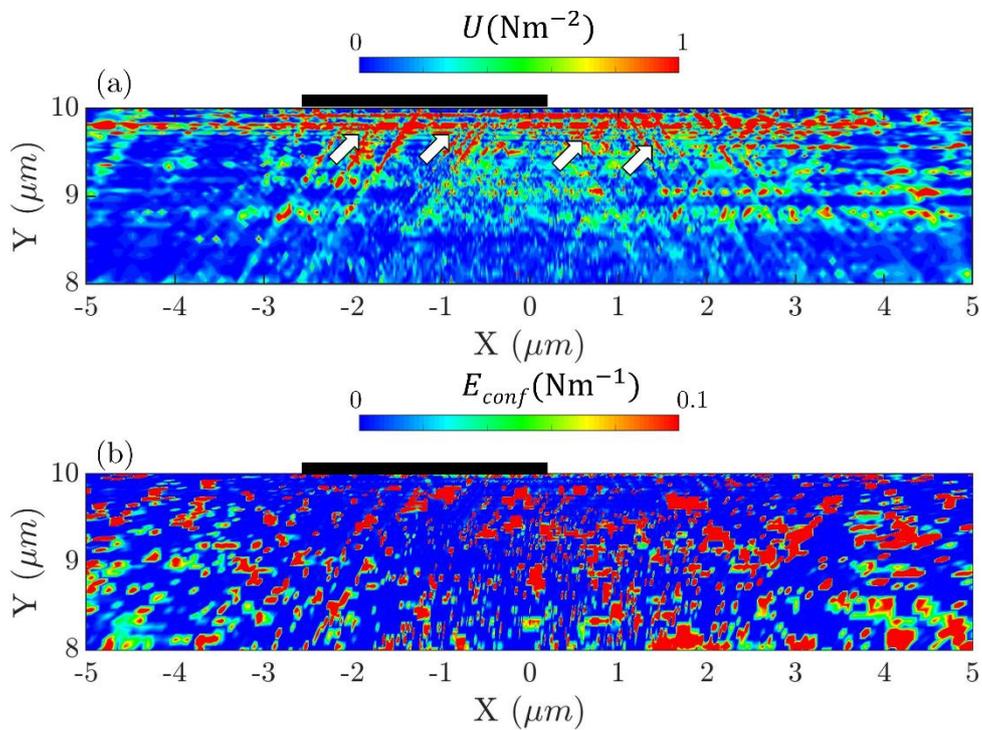

Figure A4. The comparison between (a) the plastic strain energy density and (b) the dislocation configuration energy density under the contact A=5.52μm, where full sliding is achieved.

# Reference

1. Vakis, A.I., et al., *Modeling and simulation in tribology across scales: An overview.* Tribology International, 2018. **125**: p. 169-199 DOI: http://dx.doi.org/10.1016/j.triboint.2018.02.005.

2. Bhushan, B., *Contact mechanics of rough surfaces in tribology: multiple asperity contact.* Tribology Letters, 1998. **4**(1): p. 1-35 DOI: http://dx.doi.org/10.1023/A:1019186601445.

3. Bowden, F.P., F.P. Bowden, and D. Tabor, *The friction and lubrication of solids*. Vol. 1. 2001: Oxford university press.

4. Rigney, D.A., *Large strains associated with sliding contact of metals.* Materials Research Innovations, 2016. **1**(4): p. 231-234 DOI: http://dx.doi.org/10.1007/s100190050046.

5. Bhushan, B. and M. Nosonovsky, *Scale effects in friction using strain gradient plasticity and dislocation-assisted sliding (microslip).* Acta Materialia, 2003. **51**(14): p. 4331-4345 DOI: http://dx.doi.org/10.1016/S1359-6454(03)00261-1.

6. Polonsky, I.A. and L.M. Keer, *Simulation of microscopic elastic-plastic contacts by using discrete dislocations.* Proceedings of the Royal Society a-Mathematical Physical and Engineering Sciences, 1996. **452**(1953): p. 2173-2194 DOI: http://dx.doi.org/10.1098/rspa.1996.0115.

7. Hurtado, J.A. and K.S. Kim, *Scale effects in friction of single-asperity contacts. II. Multiple-dislocation-cooperated slip.* Proceedings of the Royal Society a-Mathematical Physical and Engineering Sciences, 1999. **455**(1989): p. 3385-3400 DOI: http://dx.doi.org/10.1098/rspa.1999.0456.

8. Deshpande, V.S., A. Needleman, and E. Van der Giessen, *Discrete dislocation plasticity analysis of static friction.* Acta Materialia, 2004. **52**(10): p. 3135-3149 DOI: http://dx.doi.org/10.1016/j.actamat.2004.03.018.

9. Li, Q.Y. and K.S. Kim, *Micromechanics of friction: effects of nanometre-scale roughness.* Proceedings of the Royal Society a-Mathematical Physical and Engineering Sciences, 2008. **464**(2093): p. 1319-1343 DOI: http://dx.doi.org/10.1098/rspa.2007.0364.

10. Luan, B. and M.O. Robbins, *The breakdown of continuum models for mechanical contacts.* Nature, 2005. **435**(7044): p. 929-32 DOI: http://dx.doi.org/10.1038/nature03700.

11. Gagel, J., D. Weygand, and P. Gumbsch, *Discrete Dislocation Dynamics simulations of dislocation transport during sliding.* Acta Materialia, 2018. **156**: p. 215-227 DOI: http://dx.doi.org/10.1016/j.actamat.2018.06.002.

12. Komanduri, R., N. Chandrasekaran, and L.M. Raff, *Molecular dynamics simulation of atomic-scale friction.* Physical Review B, 2000. **61**(20): p. 14007-14019 DOI: http://dx.doi.org/10.1103/PhysRevB.61.14007.

13. Eder, S.J., et al., *Unraveling and Mapping the Mechanisms for Near-Surface Microstructure Evolution in CuNi Alloys under Sliding.* ACS Appl Mater Interfaces, 2020. **12**(28): p. 32197-32208 DOI: http://dx.doi.org/10.1021/acsami.0c09302.

14. Pan, Z. and T.J. Rupert, *Mechanisms of near-surface structural evolution in nanocrystalline materials during sliding contact.* Physical Review Materials, 2017. **1**(4): p. 043602 DOI: http://dx.doi.org/10.1103/PhysRevMaterials.1.043602.

15. Li, A. and I. Szlufarska, *How grain size controls friction and wear in nanocrystalline metals.* Physical Review B, 2015. **92**(7): p. 075418 DOI: http://dx.doi.org/10.1103/PhysRevB.92.075418.

16. Eder, S.J., et al., *Effect of Temperature on the Deformation Behavior of Copper Nickel Alloys under Sliding.* Materials, 2021. **14**(1): p. 60 DOI: http://dx.doi.org/10.3390/ma14010060.

17. Ghodrati, M., M. Ahmadian, and R. Mirzaeifar, *Three-dimensional study of rolling contact fatigue using crystal plasticity and cohesive zone method.* International Journal of Fatigue, 2019. **128** DOI: http://dx.doi.org/10.1016/j.ijfatigue.2019.105208.

18. Barzdajn, B., et al., *A crystal plasticity assessment of normally-loaded sliding contact in rough surfaces and galling.* Journal of the Mechanics and Physics of Solids, 2018. **121**: p. 517-542 DOI: http://dx.doi.org/10.1016/j.jmps.2018.08.004.

19. Godet, M., *Third-bodies in tribology.* Wear, 1990. **136**(1): p. 29-45.

20. Luo, Z.P., G.P. Zhang, and R. Schwaiger, *Microstructural vortex formation during cyclic sliding of Cu/Au multilayers.* Scripta Materialia, 2015. **107**: p. 67-70 DOI: http://dx.doi.org/10.1016/j.scriptamat.2015.05.022.

21. Liu, Z., et al., *Tribological performance and microstructural evolution of α-brass alloys as a function of zinc concentration.* Friction, 2020. **8**(6): p. 1117-1136 DOI: http://dx.doi.org/10.1007/s40544-019-0345-8.

22. Bahshwan, M., et al., *The role of microstructure on wear mechanisms and anisotropy of additively manufactured 316L stainless steel in dry sliding.* Materials & Design, 2020. **196**: p. 109076 DOI: http://dx.doi.org/10.1016/j.matdes.2020.109076.

23. Kareer, A., et al., *Scratching the surface: Elastic rotations beneath nanoscratch and nanoindentation tests.* Acta Materialia, 2020. **200**: p. 116-126 DOI: http://dx.doi.org/10.1016/j.actamat.2020.08.051.

24. Rigney, D.A. and S. Karthikeyan, *The Evolution of Tribomaterial During Sliding: A Brief Introduction.* Tribology Letters, 2010. **39**(1): p. 3-7 DOI: http://dx.doi.org/10.1007/s11249-009-9498-3.

25. Romero, P.A., et al., *Coarse graining and localized plasticity between sliding nanocrystalline metals.* Phys Rev Lett, 2014. **113**(3): p. 036101 DOI: http://dx.doi.org/10.1103/PhysRevLett.113.036101.

26. Grützmacher, P.G., et al., *Interplay between microstructural evolution and tribo-chemistry during dry sliding of metals.* Friction, 2019. **7**(6): p. 637-650 DOI: http://dx.doi.org/10.1007/s40544-019-0259-5.

27. Grützmacher, P., C. Gachot, and S.J. Eder, *Visualization of microstructural mechanisms in nanocrystalline ferrite during grinding.* Materials & Design, 2020. **195** DOI: http://dx.doi.org/10.1016/j.matdes.2020.109053.

28. Greiner, C., et al., *Sequence of Stages in the Microstructure Evolution in Copper under Mild Reciprocating Tribological Loading.* ACS Appl Mater Interfaces, 2016. **8**(24): p. 15809-19 DOI: http://dx.doi.org/10.1021/acsami.6b04035.

29. Liu, Z.L., et al., *Stages in the tribologically-induced oxidation of high-purity copper.* Scripta Materialia, 2018. **153**: p. 114-117 DOI: http://dx.doi.org/10.1016/j.scriptamat.2018.05.008.

30. Ruebeling, F., et al., *Normal Load and Counter Body Size Influence the Initiation of Microstructural Discontinuities in Copper during Sliding.* ACS Appl Mater Interfaces, 2021. **13**(3): p. 4750-4760 DOI: http://dx.doi.org/10.1021/acsami.0c19736.

31. Greiner, C., et al., *The origin of surface microstructure evolution in sliding friction.* Scripta Materialia, 2018. **153**: p. 63-67 DOI: http://dx.doi.org/10.1016/j.scriptamat.2018.04.048.

32. Xu, Y.L., et al., *On the origin of microstructural discontinuities in sliding contacts: A discrete dislocation plasticity analysis.* International Journal of Plasticity, 2021. **138** DOI: http://dx.doi.org/10.1016/j.ijplas.2021.102942.

33. Van der Giessen, E. and A. Needleman, *Discrete Dislocation Plasticity - a Simple Planar Model.* Modelling and Simulation in Materials Science and Engineering, 1995. **3**(5): p. 689-735.

34. Deshpande, V.S., et al., *Size effects in single asperity frictional contacts.* Modelling and Simulation in Materials Science and Engineering, 2007. **15**(1): p. S97-S108 DOI: http://dx.doi.org/10.1088/0965-0393/15/1/S09.

35. Xu, Y. and D. Dini, *Capturing the hardness of coating systems across the scales.* Surface and Coatings Technology, 2020. **394**: p. 125860 DOI: http://dx.doi.org/10.1016/j.surfcoat.2020.125860.

36. Balint, D.S., et al., *Discrete dislocation plasticity analysis of the grain size dependence of the flow strength of polycrystals.* International Journal of Plasticity, 2008. **24**(12): p. 2149-2172 DOI: http://dx.doi.org/10.1016/j.ijplas.2007.08.005.

37. Lubarda, V.A., J.A. Blume, and A. Needleman, *An Analysis of Equilibrium Dislocation Distributions.* Acta Metallurgica Et Materialia, 1993. **41**(2): p. 625-642 DOI: http://dx.doi.org/10.1016/0956-7151(93)90092-7.

38. Rice, J.R., *Tensile Crack Tip Fields in Elastic Ideally Plastic Crystals.* Mechanics of Materials, 1987. **6**(4): p. 317-335 DOI: http://dx.doi.org/10.1016/0167-6636(87)90030-5.

39. Xu, Y.L., et al., *Cyclic plasticity and thermomechanical alleviation in titanium alloys.* International Journal of Plasticity, 2020. **134**: p. 102753 DOI: http://dx.doi.org/10.1016/j.ijplas.2020.102753.

40. Shan, Z.W., et al., *Mechanical annealing and source-limited deformation in submicrometre-diameter Ni crystals.* Nat Mater, 2008. **7**(2): p. 115-9 DOI: http://dx.doi.org/10.1038/nmat2085.

41. Widjaja, A., E. Van der Giessen, and A. Needleman, *Discrete dislocation modelling of submicron indentation.* Materials Science and Engineering a-Structural Materials Properties Microstructure and Processing, 2005. **400**: p. 456-459 DOI: http://dx.doi.org/10.1016/j.msea.2005.01.074.

42. Widjaja, A., A. Needleman, and E. Van der Giessen, *The effect of indenter shape on sub-micron indentation according to discrete dislocation plasticity.* Modelling and Simulation in Materials Science and Engineering, 2007. **15**(1): p. S121-S131 DOI: http://dx.doi.org/10.1088/0965-0393/15/1/S11.

43. Xu, Y., D.S. Balint, and D. Dini, *A new hardness formula incorporating the effect of source density on indentation response: A discrete dislocation plasticity analysis.* Surface & Coatings Technology, 2019. **374**: p. 763-773 DOI: http://dx.doi.org/10.1016/j.surfcoat.2019.06.045.

44. Langer, J.S., E. Bouchbinder, and T. Lookman, *Thermodynamic theory of dislocation-mediated plasticity.* Acta Materialia, 2010. **58**(10): p. 3718-3732 DOI: http://dx.doi.org/10.1016/j.actamat.2010.03.009.

45. Song, H., V.S. Deshpande, and E. Van der Giessen, *Discrete dislocation plasticity analysis of loading rate-dependent static friction.* Proc Math Phys Eng Sci, 2016. **472**(2192): p. 20150877 DOI: http://dx.doi.org/10.1098/rspa.2015.0877.

46. Balint, D.S., et al., *Discrete dislocation plasticity analysis of the wedge indentation of films.* Journal of the Mechanics and Physics of Solids, 2006. **54**(11): p. 2281-2303 DOI: http://dx.doi.org/10.1016/j.jmps.2006.07.004.

47. Uchic, M.D., et al., *Sample dimensions influence strength and crystal plasticity.* Science, 2004. **305**(5686): p. 986-9 DOI: http://dx.doi.org/10.1126/science.1098993.

48. Kysar, J.W., et al., *Experimental lower bounds on geometrically necessary dislocation density.* International Journal of Plasticity, 2010. **26**(8): p. 1097-1123 DOI: http://dx.doi.org/10.1016/j.ijplas.2010.03.009.

49. Prastiti, N.G., et al., *Discrete dislocation, crystal plasticity and experimental studies of fatigue crack nucleation in single-crystal nickel.* International Journal of Plasticity, 2020. **126**: p. 102615 DOI: http://dx.doi.org/10.1016/j.ijplas.2019.10.003.

50. Zhang, Y.H., Y.F. Gao, and L. Nicola, *Lattice rotation caused by wedge indentation of a single crystal: Dislocation dynamics compared to crystal plasticity simulations.* Journal of the Mechanics and Physics of Solids, 2014. **68**: p. 267-279 DOI: http://dx.doi.org/10.1016/j.jmps.2014.04.006.

51. Peng, B., et al., *Effect of shear stress on adhesive contact with a generalized Maugis-Dugdale cohesive zone model.* Journal of the Mechanics and Physics of Solids, 2021. **148** DOI: http://dx.doi.org/10.1016/j.jmps.2020.104275.

52. Menga, N., G. Carbone, and D. Dini, *Do uniform tangential interfacial stresses enhance adhesion?* Journal of the Mechanics and Physics of Solids, 2018. **112**: p. 145-156 DOI: http://dx.doi.org/https://doi.org/10.1016/j.jmps.2017.11.022.

53. Mergel, J.C., J. Scheibert, and R.A. Sauer, *Contact with coupled adhesion and friction: Computational framework, applications, and new insights.* Journal of the Mechanics and Physics of Solids, 2021. **146**: p. 104194 DOI: http://dx.doi.org/https://doi.org/10.1016/j.jmps.2020.104194.

54. Aghababaei, R., D.H. Warner, and J.-F. Molinari, *Critical length scale controls adhesive wear mechanisms.* Nature Communications, 2016. **7**(1): p. 11816 DOI: http://dx.doi.org/10.1038/ncomms11816.

55. Gao, Y.F., et al., *Nanoscale incipient asperity sliding and interface micro-slip assessed by the measurement of tangential contact stiffness.* Scripta Materialia, 2006. **55**(7): p. 653-656 DOI: http://dx.doi.org/10.1016/j.scriptamat.2006.05.006.

56. Hanke, S., J. Petri, and D. Johannsmann, *Partial slip in mesoscale contacts: dependence on contact size.* Phys Rev E Stat Nonlin Soft Matter Phys, 2013. **88**(3): p. 032408 DOI: http://dx.doi.org/10.1103/PhysRevE.88.032408.

57. Shen, F., W.P. Hu, and Q.C. Meng, *A damage mechanics approach to fretting fatigue life prediction with consideration of elastic-plastic damage model and wear.* Tribology International, 2015. **82**: p. 176-190 DOI: http://dx.doi.org/10.1016/j.triboint.2014.10.017.

58. Hinkle, A.R., et al., *The emergence of small-scale self-affine surface roughness from deformation.* Science Advances, 2020. **6**(7): p. eaax0847 DOI: http://dx.doi.org/10.1126/sciadv.aax0847.

59. Stinville, J.C., et al., *Direct measurements of slip irreversibility in a nickel-based superalloy using high resolution digital image correlation.* Acta Materialia, 2020. **186**: p. 172-189 DOI: http://dx.doi.org/10.1016/j.actamat.2019.12.009.

60. Miao, J.S., T.M. Pollock, and J.W. Jones, *Microstructural extremes and the transition from fatigue crack initiation to small crack growth in a polycrystalline nickel-base superalloy.* Acta Materialia, 2012. **60**(6-7): p. 2840-2854 DOI: http://dx.doi.org/10.1016/j.actamat.2012.01.049.

61. Stinville, J.C., et al., *High resolution mapping of strain localization near twin boundaries in a nickel-based superalloy.* Acta Materialia, 2015. **98**: p. 29-42 DOI: http://dx.doi.org/10.1016/j.actamat.2015.07.016.

62. Dunne, F.P.E., D. Rugg, and A. Walker, *Lengthscale-dependent, elastically anisotropic, physically-based hcp crystal plasticity: Application to cold-dwell fatigue in Ti alloys.* International Journal of Plasticity, 2007. **23**(6): p. 1061-1083 DOI: http://dx.doi.org/10.1016/j.ijplas.2006.10.013.

63. Haug, C., et al., *Early deformation mechanisms in the shear affected region underneath a copper sliding contact.* Nat Commun, 2020. **11**(1): p. 839 DOI: http://dx.doi.org/10.1038/s41467-020-14640-2.

64. Brown, A.A. and D.J. Bammann, *Validation of a model for static and dynamic recrystallization in metals.* International Journal of Plasticity, 2012. **32-33**: p. 17-35 DOI: http://dx.doi.org/10.1016/j.ijplas.2011.12.006.

65. Xu, Y., *A non-local methodology for geometrically necessary dislocations and application to crack tips.* International Journal of Plasticity, 2021: p. 102970 DOI: http://dx.doi.org/10.1016/j.ijplas.2021.102970.

66. Bergsmo, A., et al., *Twin boundary fatigue crack nucleation in a polycrystalline Nickel superalloy containing non-metallic inclusions.* Journal of the Mechanics and Physics of Solids, 2022. **160** DOI: http://dx.doi.org/10.1016/j.jmps.2022.104785.

67. Cheong, K.S., E.P. Busso, and A. Arsenlis, *A study of microstructural length scale effects on the behaviour of FCC polycrystals using strain gradient concepts.* International Journal of Plasticity, 2005. **21**(9): p. 1797-1814 DOI: http://dx.doi.org/10.1016/j.ijplas.2004.11.001.

68. Arsenlis, A. and D.M. Parks, *Crystallographic aspects of geometrically-necessary and statistically-stored dislocation density.* Acta Materialia, 1999. **47**(5): p. 1597-1611 DOI: http://dx.doi.org/10.1016/S1359-6454(99)00020-8.

69. Kiener, D., et al., *Work hardening in micropillar compression: In situ experiments and modeling.* Acta Materialia, 2011. **59**(10): p. 3825-3840 DOI: http://dx.doi.org/10.1016/j.actamat.2011.03.003.